\documentclass{article}

\usepackage{arxiv}

\usepackage[utf8]{inputenc} 
\usepackage[T1]{fontenc}    
\usepackage{hyperref}       
\usepackage{url}            
\usepackage{booktabs}       
\usepackage{amsfonts}       
\usepackage{nicefrac}       
\usepackage{microtype}      
\usepackage{graphicx}
\usepackage{natbib}
\usepackage{doi}
\usepackage{placeins}   
\usepackage{tikz}
\usetikzlibrary{shapes.geometric,shapes.misc,arrows,positioning,fit,backgrounds}

\title{Bias-Aware AI Chatbot for Engineering Advising at the University of Maryland A. James Clark School of Engineering}

\author{
	Prarthana P.~Kartholy\thanks{These authors contributed equally.} \\
	University of Maryland, College Park\\
	A. James Clark School of Engineering\\
	ESTEEM/SER-Quest \\
	\texttt{pillapayi1@gmail.com} \\
	\And
	Thandi M.~Labor\footnotemark[1] \\
	University of Maryland, College Park\\
	A. James Clark School of Engineering\\
	ESTEEM/SER-Quest \\
	\texttt{Thandi.labor@gmail.com} \\
	\And
	Neil N.~Panchal\footnotemark[1] \\
	University of Maryland, College Park\\
	A. James Clark School of Engineering\\
	ESTEEM/SER-Quest \\
	\texttt{neilpanchal08@gmail.com} \\
	\And
	Sean H.~Wang\footnotemark[1] \\
	University of Maryland, College Park\\
	A. James Clark School of Engineering\\
	ESTEEM/SER-Quest \\
	\texttt{seanwangpiano@gmail.com} \\
	\And
	Hillary N.~Owusu\thanks{To whom correspondence should be addressed.} \\
	University of Maryland, College Park\\
	A. James Clark School of Engineering\\
	ESTEEM/SER-Quest \\
	\texttt{hnyowusu@umd.edu} \\
}

\renewcommand{\shorttitle}{Bias-Aware AI Chatbot for Engineering Advising}

\hypersetup{
pdftitle={Bias-Aware AI Chatbot for Engineering Advising at the University of Maryland A. James Clark School of Engineering},
pdfsubject={cs.AI, cs.CY},
pdfauthor={Prarthana P.~Kartholy, Thandi M.~Labor, Neil N.~Panchal, Sean H.~Wang, Hillary N.~Owusu},
pdfkeywords={AI chatbot, bias mitigation, academic advising, engineering education, algorithmic fairness},
}

\begin{document}
\maketitle

\begin{abstract}
	Selecting a college major is a difficult decision for many incoming freshmen. Traditional academic advising is often hindered by long wait times, intimidating environments, and limited personalization. AI Chatbots present an opportunity to address these challenges. However, AI systems also have the potential to generate biased responses, prejudices related to race, gender, socioeconomic status, and disability. These biases risk turning away potential students and undermining reliability of AI systems. This study aims to develop a University of Maryland (UMD) A James Clark Engineering Program-specific AI chatbot. Our research team analyzed and mitigated potential biases in the responses. Through testing the chatbot on diverse student queries, the responses are scored on metrics of accuracy, relevance, personalization, and bias presence. The results demonstrate that with careful prompt engineering and bias mitigation strategies, AI chatbots can provide high-quality, unbiased academic advising support, achieving mean scores of 9.76 for accuracy, 9.56 for relevance, and 9.60 for personalization with no stereotypical biases found in the sample data. However, due to the small sample size and limited timeframe, our AI model may not fully reflect the nuances of student queries in engineering academic advising. Regardless, these findings will inform best practices for building ethical AI systems in higher education, offering tools to complement traditional advising and address the inequities faced by many underrepresented and first-generation college students.
\end{abstract}


\section{Introduction}

\subsection{Background and Context}

The transition from secondary to higher education requires students to make crucial decisions regarding their areas of studies. Specifically, students matriculating into the A. James Clark School of Engineering must choose from 22 programs. UMD's traditional college advising system provides necessary guidance to students during these decisions. Yet the current system exhibits notable limitations: appointment wait times, intimidating formal environments, and reduced information retention, all of which create barriers to academic guidance \citep{fox2022student}.

Artificial intelligence (AI) chatbots have proven complementary tools to address the accessibility limitations in traditional advising, academic or otherwise. These adaptive systems offer distinct advantages including continuous availability, minimizing social barriers, and standardized information delivery \citep{hasal2021chatbots}. Implementing a chatbot alternative suggests a potential for enhanced accessibility to academic advising services, particularly for students who may be reluctant to engage with traditional advising.

\subsection{Problem Statement}

Nevertheless, the use of AI infrastructure in educational institutions for advising contexts presents significant challenges and concerns regarding algorithmic bias and its potential to exacerbate existing educational inequities. Empirical evidence indicates that Large Language Models (LLMs) tend to rely on stereotypical assumptions and demonstrate varying response quality when users directly or indirectly reveal identity information \citep{smith2024sorry}. These documented disparities raise critical questions about equitable implementation of AI-based advising tools and highlight the need for the integration of bias mitigation strategies in AI based tools.

This study investigates how to implement a bias-aware AI chatbot for engineering advising at the University of Maryland A. James Clark School of Engineering, to address algorithmic bias and its role in reinforcing educational inequities.

\subsection{Literature Review}

Research on AI chatbots in educational contexts reveals significant challenges related to algorithmic bias, user interaction dynamics, and security requirements that must be addressed for ethical implementation in engineering advising.

\subsubsection{Algorithmic Bias in Educational AI}

Multiple studies document systematic biases in large language models used for educational recommendations. Smith and the Institute for Advancing Computing Education (2024) found that ChatGPT exhibited statistically significant racial bias, being more likely to recommend STEM majors to Hispanic and Asian students compared to Black or White students when presented with identical academic profiles. Claude AI similarly recommended colleges with lower SAT scores and salaries to Black students. Kantharuban et al. (2024) demonstrated that even subtle identity cues influence recommendations regarding college prestige and type, with inconsistent treatment across demographic groups.

Bias extends beyond race to gender and disability. Parra et al. (2022) documented AI systems that disproportionately showed high-paying positions to men, while Urbina et al. (2024) found that ChatGPT and Gemini consistently used negative language when describing people with disabilities, systematically underestimating their capabilities.

\subsubsection{Human-AI Interaction Complexity}

Wang et al. (2023) revealed a concerning paradox through testing with over 200 college students: users often preferred biased (gender-aware) systems over unbiased alternatives, even when unbiased systems demonstrated superior accuracy and fairness. Participants actively rejected recommendations for careers dominated by the opposite gender, despite alignment with their interests, suggesting that effective bias mitigation must address both algorithmic and human cognitive biases.

\subsubsection{Ethical Data and Security Framework}

Daly et al. (2021) argue that ethical AI requires ``Good Data'' practices that avoid reinforcing social hierarchies, emphasizing that bias mitigation requires fundamental attention to data sourcing and curation rather than merely technical adjustments.

Security considerations are equally critical. Hasal et al. (2021) identify two key domains: message transfer security and backend data storage, emphasizing authentication protocols, encryption, and GDPR compliance. Yang et al. (2023) found that many chatbots lack adequate privacy policies, while Araya (2022) and Miller (2024) highlight how the economic value of data creates incentives for sophisticated cyberattacks, necessitating robust user consent mechanisms and regulatory frameworks.

The literature establishes that bias-aware AI chatbot development requires integrated approaches addressing technical bias detection, ethical data practices, security measures, and user education to prevent perpetuation of educational inequities.

\subsection{Research Question}

How can we implement an AI chatbot for engineering advising at the University of Maryland A. James Clark School of Engineering that mitigates algorithmic bias while maintaining high accuracy, relevance, and personalization of responses?

\subsection{Hypothesis}

By utilizing systematic prompt engineering, comprehensive data curation, and bias detection system, we can develop an AI chatbot that provides high-quality academic advising with minimal detectable bias across diverse student populations.

\section{Methodology}
\label{sec:methodology}

In our methods, we applied the following software tools to process and analyze data and create a prototype of our AI Chatbot:
\begin{itemize}
	\item \textbf{Python 3}: Our main coding language for our AI chatbot and data analysis program.
	\item \textbf{Regular Expressions}: To identify prompt categories and potential biases.
	\item \textbf{Pandas and Matplotlib}: Python libraries used for data analysis and visualization.
	\item \textbf{VSCode}: As the coding environment for the AI Chatbot.
	\item \textbf{Jupyter Notebook}: As the coding environment for the data analysis program.
	\item \textbf{Mistral Medium AI}: An open source LLM for the base model of our chatbot.
	\item \textbf{Gradio}: An open source python library for creating the UI Interface.
	\item \textbf{Beautiful Soup and Scrapy}: Open-source libraries for web scraping.
\end{itemize}

Here is the chronological methodology:

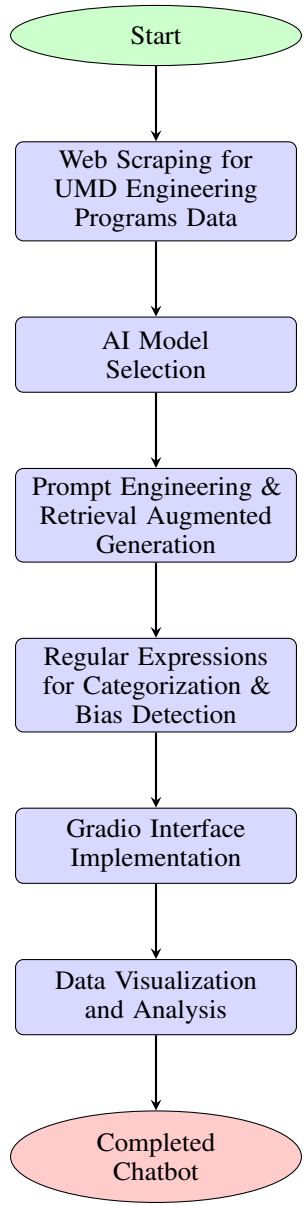
\begin{figure}[h!]
	\centering
	\begin{tikzpicture}[
		node distance=1cm,
		start/.style={ellipse, draw, fill=green!20, text width=2.5cm, text centered, minimum height=0.8cm},
		process/.style={rectangle, draw, fill=blue!15, text width=3.5cm, text centered, minimum height=1cm, rounded corners=3pt},
		decision/.style={diamond, draw, fill=yellow!20, text width=2.5cm, text centered, aspect=2, inner sep=0pt},
		end/.style={ellipse, draw, fill=red!20, text width=2.5cm, text centered, minimum height=0.8cm},
		arrow/.style={thick, ->, >=stealth}
	]
	
	\node[start] (start) {Start};
	\node[process, below=of start] (data) {Web Scraping for\\UMD Engineering\\Programs Data};
	\node[process, below=of data] (model) {AI Model\\Selection};
	\node[process, below=of model] (develop) {Prompt Engineering \&\\Retrieval Augmented\\Generation};
	\node[process, below=of develop] (regex) {Regular Expressions\\for Categorization \&\\Bias Detection};
	\node[process, below=of regex] (gradio) {Gradio Interface\\Implementation};
	\node[process, below=of gradio] (analysis) {Data Visualization\\and Analysis};
	\node[end, below=of analysis] (end) {Completed\\Chatbot};
	
	\draw[arrow] (start) -- (data);
	\draw[arrow] (data) -- (model);
	\draw[arrow] (model) -- (develop);
	\draw[arrow] (develop) -- (regex);
	\draw[arrow] (regex) -- (gradio);
	\draw[arrow] (gradio) -- (analysis);
	\draw[arrow] (analysis) -- (end);
	
	\end{tikzpicture}
	\caption{Flowchart demonstrating workflow methodology for developing the bias-aware AI chatbot following the methodology subsection structure.}
	\label{fig:workflow}
\end{figure}

\FloatBarrier

\subsection{Web Scraping for UMD Engineering Programs Data}

A data collection process utilizing spider, an automated website scrapping program, was implemented to gather extensive information about UMD's engineering programs: program descriptions, course requirements, prerequisites, career pathways, faculty information, etc. The scraping process also incorporated Beautiful Soup and Scrapy, open-source Python libraries specifically designed for web scraping applications, to parse HTML content and extract relevant information from multiple university web pages.

The extracted data was structured and stored in JSON format to facilitate efficient retrieval by the chatbot. Data validation procedures were also implemented to ensure accuracy and completeness, including cross validation across multiple sources from UMD and removing outdated or inconsistent data. Our team then manually enhanced the JSON database by adding more keywords and descriptive tags, expanding the breadth of information for all engineering majors offered at UMD, including specialized tracks, interdisciplinary programs, and graduate pathway options.

\subsection{AI Model Selection}

The selection of the underlying AI model was based on a detailed evaluation of available LLMs. After comparative analysis of multiple models, including OpenAI's GPT-3.5, Meta's LLaMA 2, and Cohere's Command R, particular attention was given to more open-source alternatives. Mistral Medium AI was selected as the primary language model due to its strong performance across several criteria relevant to our use case. Specifically, this open-source LLM demonstrated optimal performance for educational applications while providing transparency in bias mitigation approaches. We set the Top-P at 1.0 and the temperature at 0.7. Top-P (nucleus sampling) controls the diversity of the model's output by limiting the probability mass considered during generation, while temperature adjusts the randomness of the model's responses (higher value 1 = more varied outputs; lower value 0 = more deterministic outputs).

As our research progressed, we found that a lower temperature produced more consistent and reliable responses, which was crucial for data analysis and bias detection. Therefore, in later iterations of our evaluation, we set temperature to 0.0. This shift in temperature ensured that the model generated the most probable and standardized responses for each prompt, minimizing the variability. Moreover, we conducted three distinct rounds of data collection using these finalized parameters, which allowed us to more accurately assess the model's behavior and response quality.

\subsection{Prompt Engineering and Retrieval Augmented Generation}

A framework of nine distinct prompt categories was developed to enhance the AI model's performance. We created a base prompt, which was added to the user input before submitting to the LLM. The base prompt contains instructions and context for the Chatbot, such as taking the role of an academic advisor and avoiding stereotypical phrases like "usually" and "typically". Additionally, the base prompt uses a basic Retrieval Augmented Generation (RAG) system, in which the JSON file was read and processed into plain text information, then appended to the base prompt.

\subsection{Regular Expressions for Categorization and Bias Detection}

Regex expressions were created for performing lexical search on the prompt for category identification. The prompts were labeled with one or more of nine categories: affirmative, negative, vague, specific, identity\_disclosure, time\_oriented, interest\_broad, interest\_narrow, and general. These categories are later used for category-specific analysis of the chatbot.

Additionally, regular expressions were used for the bias identification process, in addition to a manual qualitative analysis. For example, the regular expressions searched for stereotypical phrases like "People from your background typically..." and "[Race/Gender] usually ..." to help detect and remove potential biases.

\subsection{Gradio Interface Implementation}

The Gradio library was used for creating a User Interface with the chatbot's input and output. Our developer model features the Evaluate Responses tab, which allows us to score the responses and submit them to a csv file. Responses are scored out of 10 on accuracy, relevance, and personalization. The Analysis Dashboard tab features an export button, allowing us to save the csv file for later use.

\begin{figure}[h!]
	\centering
	\includegraphics[width=0.8\textwidth]{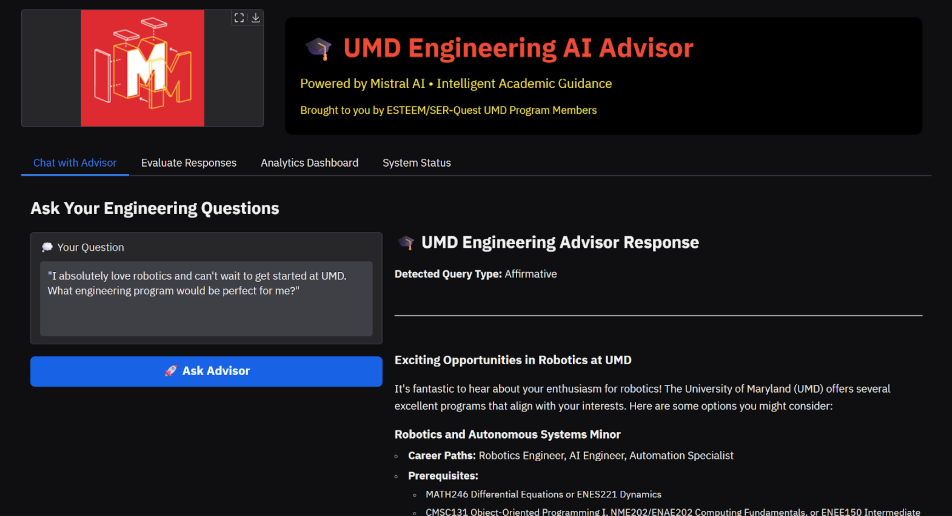}
	\caption{Gradio interface V1 of Chatbot showing the user interaction panel and evaluation tools.}
	\label{fig:gradio}
\end{figure}

\FloatBarrier

\subsection{Data Visualization and Analysis}

A jupyter notebook was created to visualize the CSV files mentioned above. For the three Temperature = 0.0 iterations, the jupyter notebook averaged scores in corresponding data cells across the 3 csv files prior to data visualization. The notebook utilizes Pandas to retrieve means and standard deviations of the accuracy, relevance and personalization. We used matplotlib to generate bar charts and histograms for the score distributions.

\section{Results}
\label{sec:results}

To assess the quality of the AI chatbot's advising responses, a manual review was conducted on a sample of 75 prompts by our research team. Each response was scored on a scale from 1 to 10 across three dimensions: accuracy, relevance, and personalization. Responses were categorized into 9 distinct prompt categories: Affirmative, Negative, Identity disclosure, Vague, Specific, Time Oriented, Broad Interest, Narrow Interest, and General.

\subsection{Temperature = 0.7 Results}

\begin{figure}[h!]
	\centering
	\includegraphics[width=\textwidth]{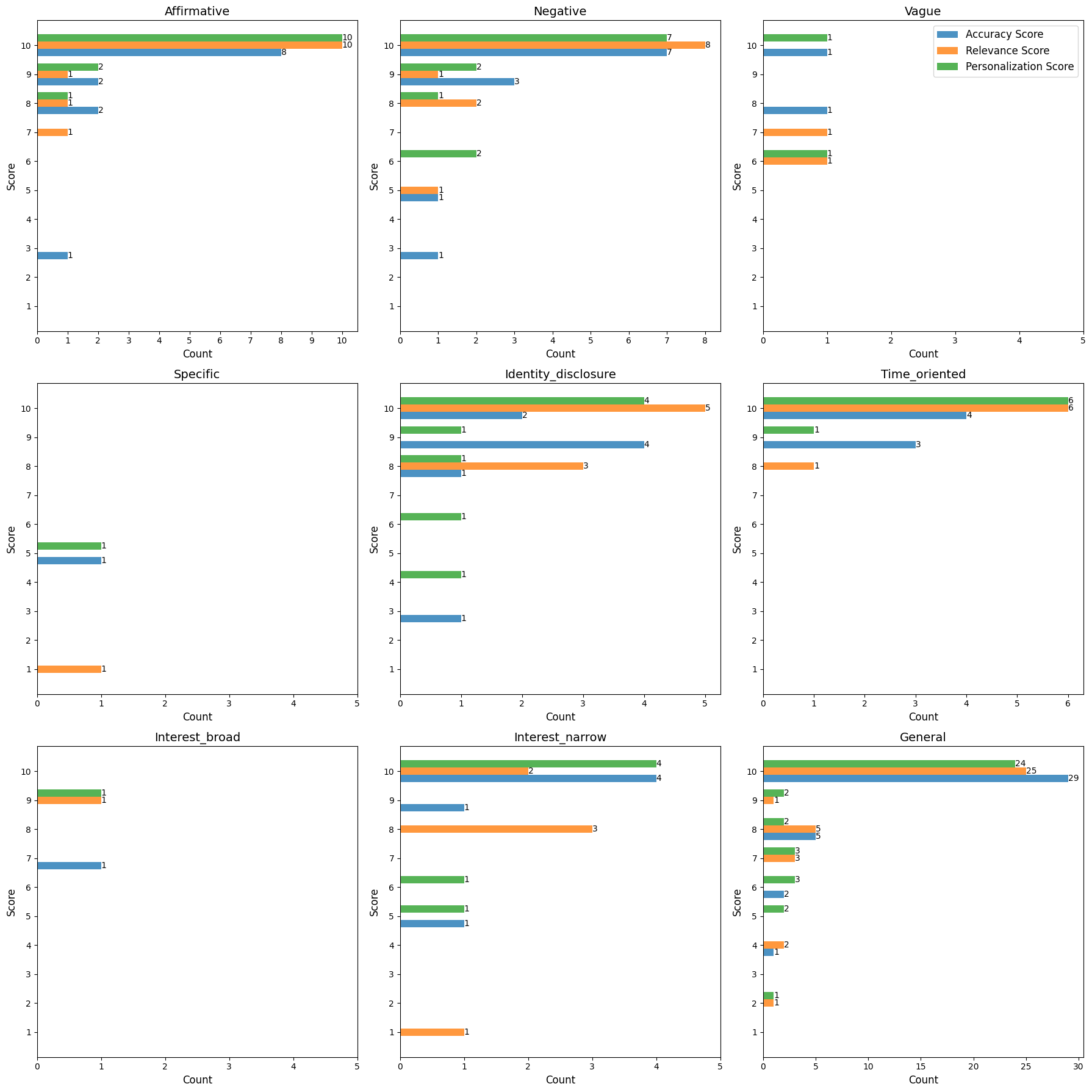}
	\caption{Score Frequency Distributions by Prompt Category for Temperature = 0.7}
	\label{fig:temp07_distribution}
\end{figure}

For each category, a bar chart was created to visualize the frequency distribution of each of the 3 scores, as shown in Figure~\ref{fig:temp07_distribution}.

\FloatBarrier
\begin{table}[h!]
	\caption{Temperature = 0.7 - Mean and Standard Deviation of Scores by Prompt Category}
	\centering
	\small
	\begin{tabular}{lccccccc}
		\toprule
		Category & Count & \multicolumn{2}{c}{Accuracy} & \multicolumn{2}{c}{Relevance} & \multicolumn{2}{c}{Personalization} \\
		& & Mean & SD & Mean & SD & Mean & SD \\
		\midrule
		Affirmative & 13 & 9.00 & 1.96 & 9.54 & 0.97 & 9.69 & 0.63 \\
		Negative & 12 & 8.75 & 2.30 & 9.17 & 1.53 & 9.00 & 1.54 \\
		Identity Disclosure & 8 & 8.38 & 2.26 & 9.25 & 1.04 & 8.38 & 2.26 \\
		Time Oriented & 7 & 9.57 & 0.53 & 9.71 & 0.76 & 9.86 & 0.38 \\
		Interest Narrow & 6 & 9.00 & 2.00 & 7.50 & 3.33 & 8.50 & 2.35 \\
		General & 37 & 9.35 & 1.42 & 8.92 & 1.98 & 8.78 & 2.00 \\
		\midrule
		Overall & 75 & 9.09 & 1.68 & 8.96 & 1.91 & 8.97 & 1.81 \\
		\bottomrule
	\end{tabular}
	\label{tab:temp07_results}
\end{table}

Note: Counts do not sum to 75 due to overlapping category assignments. Some categories are excluded due to insufficient data.

\FloatBarrier

Overall, the model performed well on Affirmative, Negative, Identity-Disclosure, and General responses, having a majority of responses scored 8-10 across all three evaluation criteria. The overall means of the 3 scoring dimensions were 9.09, 8.96, and 8.97.

A qualitative review of some responses further highlighted these trends. High-scoring responses tended to include program-specific suggestions aligned with the student's expressed interests. For example:

\textbf{Prompt:} "I absolutely love robotics and can't wait to get started at UMD. What engineering program would be perfect for me?"

\textbf{Response:} "That's fantastic to hear about your passion for robotics! The University of Maryland (UMD) offers several programs that align well with your interests..."

\textbf{Scores:} 10 (accuracy), 10 (relevance), 10 (personalization)

In contrast, some low-scoring responses contained factual inaccuracies:

\textbf{Prompt:} "I'm super excited about studying biomedical engineering at UMD! What courses should I take in my first semester?"

\textbf{Response:} "That's fantastic! Biomedical engineering is an exciting and rapidly growing field. While UMD does not offer a dedicated biomedical..."

\textbf{Scores:} 3 (accuracy), 7 (relevance), 8 (personalization)

Note: UMD does in fact offer a biomedical engineering major.

In addition to the quantitative scoring, bias detection was performed using regular expression (regex) pattern matching on the chatbot's responses. Regex filtering ensures the chatbot excludes UMD's demographic and population data from the recommendation process. The regex targeted demographics references, followed by stereotype-related qualifiers like "typically" and "usually," and comparative terms such as "better" or "worse". The bias presence rate was 0\%.

\subsection{Temperature = 0.0 Results}

\begin{figure}[h!]
	\centering
	\includegraphics[width=\textwidth]{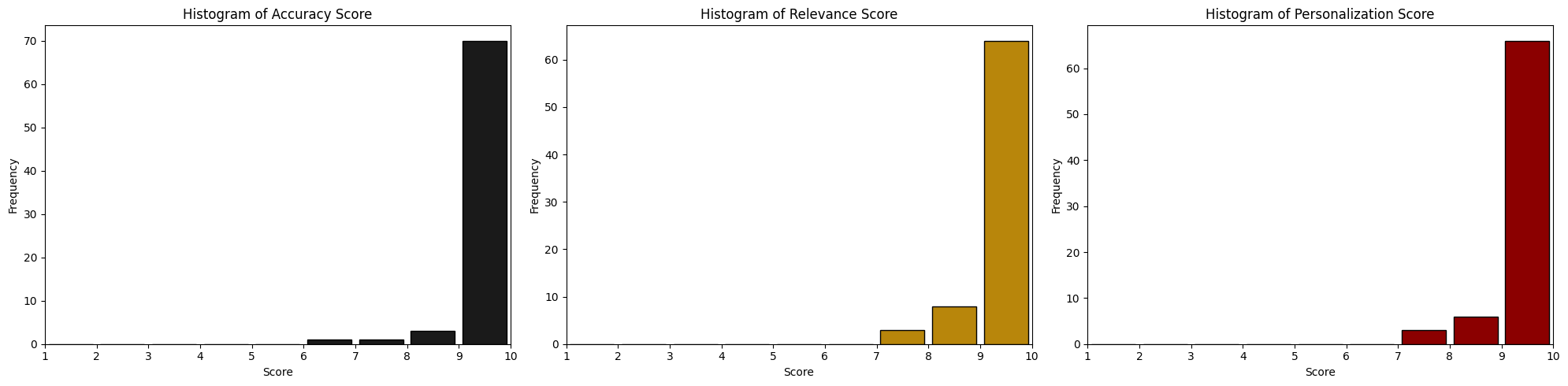}
	\caption{Histogram of scores for 75 prompts at Temperature = 0.0}
	\label{fig:temp00_histogram}
\end{figure}
\FloatBarrier
\begin{table}[h!]
	\caption{Temperature = 0.0 - Mean and Standard Deviation of Scores by Prompt Category}
	\centering
	\small
	\begin{tabular}{lccccccc}
		\toprule
		Category & Count & \multicolumn{2}{c}{Accuracy} & \multicolumn{2}{c}{Relevance} & \multicolumn{2}{c}{Personalization} \\
		& & Mean & SD & Mean & SD & Mean & SD \\
		\midrule
		Affirmative & 13.0 & 9.79 & 0.65 & 9.79 & 0.48 & 9.82 & 0.65 \\
		Negative & 12.0 & 9.83 & 0.41 & 9.64 & 0.64 & 9.72 & 0.68 \\
		Identity Disclosure & 8.0 & 9.71 & 0.55 & 9.63 & 0.63 & 9.75 & 0.39 \\
		Time Oriented & 7.0 & 9.95 & 0.13 & 9.52 & 0.81 & 9.38 & 0.97 \\
		Interest Narrow & 6.0 & 9.78 & 0.54 & 9.33 & 1.05 & 9.56 & 0.81 \\
		General & 37.0 & 9.69 & 0.75 & 9.45 & 0.72 & 9.53 & 0.65 \\
		\midrule
		Overall & 75.0 & 9.76 & 0.63 & 9.56 & 0.68 & 9.60 & 0.67 \\
		\bottomrule
	\end{tabular}
	\label{tab:temp00_results}
\end{table}

Note: Counts do not sum to 75 due to overlapping category assignments. Some categories are excluded due to insufficient data.

\FloatBarrier

The temperature adjustment from 0.7 to 0.0 resulted in significant improvements across all metrics. The mean of accuracy improved from 9.09 to 9.76, relevance from 8.96 to 9.56, and personalization from 8.97 to 9.60. Additionally, the standard deviations decreased substantially (accuracy SD from 1.68 to 0.63, relevance SD from 1.91 to 0.68, personalization SD from 1.81 to 0.67), indicating more consistent performance across different prompt types. The bias presence rate remained at 0.0\%.

\section{Discussion}
\label{sec:discussion}

\subsection{Effect of Temperature on Model Performance}

The results of this study provide significant insights into the development and implementation of bias-aware AI chatbots for engineering advising contexts. Our findings demonstrate that with careful design considerations, AI chatbots can achieve high performance metrics while maintaining ethical standards.

Previously, with Temperature = 0.7, time oriented responses resulted in the strongest responses across all three evaluation dimensions (accuracy: 9.57, relevance: 9.71, personalization: 9.86), demonstrating how specific temporal context improves response quality. This finding aligns with existing research on AI performance, where contextual specificity enhances response quality \citep{wang2023biased}. Identity disclosure responses had weaker responses (8.38, 9.25, 8.38), demonstrating how the AI struggled to find information tailored to specific identities.

When temperature was lowered to 0.0, all categories had increased scoring means and decreased scoring standard deviations; there was an overall increase in consistent high quality responses. This finding supports the hypothesis that engineering the model for deterministic responses is preferable in educational advising contexts, where consistency and reliability are paramount. The reduction in standard deviations (accuracy: 1.68 to 0.63, relevance: 1.91 to 0.68, personalization: 1.81 to 0.67) indicates that lower temperature settings provide more predictable performance, which is essential for deployment in real-world advising scenarios.

\subsection{Bias Mitigation Success}

The 0\% bias presence rate across all tested prompts represents a significant achievement in AI ethics implementation. This result demonstrates that proactive bias mitigation strategies, including careful prompt engineering and the exclusion of stereotypical language patterns, can effectively prevent biased outputs. Our approach of explicitly instructing the model to avoid terms like "typically" and "usually" when discussing demographic groups proved successful, aligning with recommendations from recent bias mitigation research \citep{kantharuban2024stereotype}.

While this method provides preliminary insight into bias presence, it is limited by the specificity of the regex patterns and the size of the dataset. Nonetheless, the absence of detectable bias patterns, achieved through targeted prompt engineering, represents an important first step toward advancing the understanding of bias in chatbot-based academic advising.

\subsection{Limitations and Considerations}

Admittedly, there are several limitations to the interpretation of our results. The relatively small sample size (75 prompts) limits statistical power and may not capture the full range of potential bias scenarios. Some of the bar charts and histograms only contained one or two data points. Our regex-based bias detection, while systematic, may miss subtle forms of bias that require more sophisticated natural language processing techniques. Additionally, the manual scoring process, despite efforts at objectivity, introduces potential evaluator bias.

The time constraints mentioned in our methodology significantly impacted both the development process and evaluation scope. We lacked the time necessary for more extensive model selection, web scraping, coding, and iterations. More extensive testing with diverse student populations and expanded prompt categories would strengthen the validity of our findings. Future iterations should incorporate more advanced bias detection tools and larger-scale user testing.

\section{Conclusion}
\label{sec:conclusion}

This study set out to develop and evaluate a bias-aware AI chatbot designed to support academic advising within the University of Maryland's A. James Clark School of Engineering. The project focused on minimizing bias while ensuring that the chatbot delivers accurate, relevant, and personalized guidance to a diverse student population. Over the course of development, the chatbot was systematically engineered through prompt design, data refinement, and the integration of bias detection tools to support equitable advising outcomes.

Our hypothesis, that an AI chatbot can provide high-quality advising with minimal bias across different student demographics, was supported. The model achieved strong performance metrics, with an average accuracy score of 9.76, relevance score of 9.56, and personalization score of 9.60. Notably, bias detection tools reported a 0\% bias rate across all tested prompts, underscoring the effectiveness of the bias mitigation strategies employed throughout development.

Looking forward, future improvements will prioritize enhancing user engagement to strengthen interaction quality and overall effectiveness. Incorporating Retrieval-Augmented Generation (RAG) with Vector (Semantic) Search will be another key step toward improving the depth and contextual accuracy of responses. Continued focus on fairness and inclusivity will remain central to ensuring that the system serves all students equitably across the Clark School of Engineering.

\bibliographystyle{unsrtnat}
\bibliography{references}

\appendix

\section{Stereotype Patterns in VSCode}
\label{app:stereotype_patterns}

\begin{verbatim}
stereotype_patterns = [
    (
        r"(?:girls|women|females?)\s+(?:are\s+)?" +
        r"(?:typically|usually|often|naturally)\s+" +
        r"(?:better|worse|good|bad)", 
        "Gender stereotype"
    ),
    (
        r"(?:boys|men|males?)\s+(?:are\s+)?" +
        r"(?:typically|usually|often|naturally)\s+" +
        r"(?:better|worse|good|bad)", 
        "Gender stereotype"
    ),
    (
        r"students?\s+like\s+you\s+" +
        r"(?:usually|typically|often)",
        "Identity-based assumption"
    ),
    (
        r"people\s+from\s+your\s+background", 
        "Background assumption"
    )
]
\end{verbatim}

\section{Commercial Model Interface Design}
\label{app:commercial_model}

\begin{figure}[h!]
	\centering
	\includegraphics[width=0.8\textwidth]{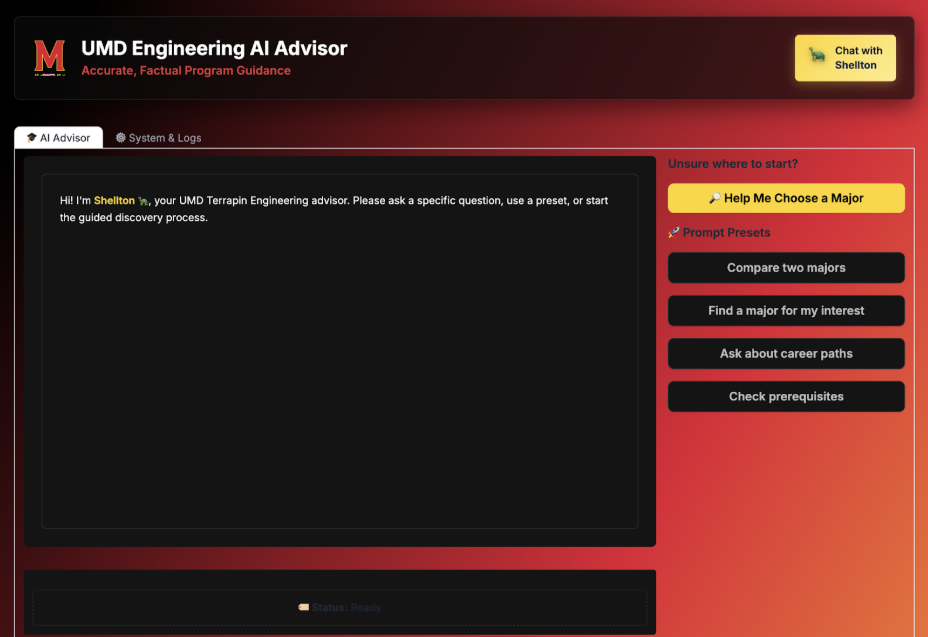}
	\caption{V2 Commercial Model interface design for the AI chatbot showing enhanced user experience features and improved layout.}
	\label{fig:commercial_v2}
\end{figure}
\FloatBarrier

\section{Temperature = 0.0 Score Distributions by Category}
\label{app:temp00_distributions}

\begin{figure}[h!]
	\centering
	\includegraphics[width=0.8\textwidth]{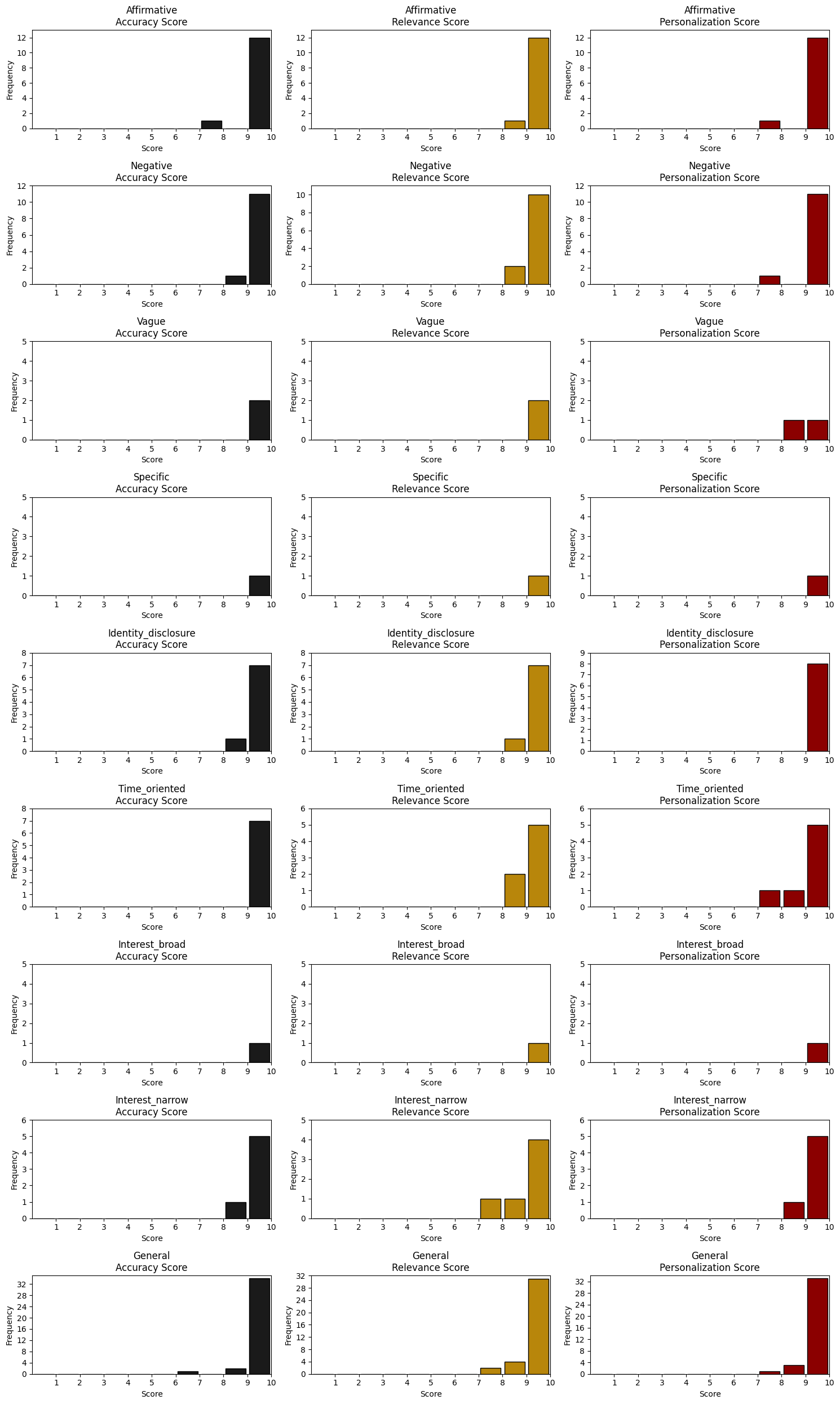}
	\caption{Detailed score distributions by prompt category for Temperature = 0.0, showing accuracy, relevance, and personalization metrics across all nine prompt categories.}
	\label{fig:temp00_categories}
\end{figure}
\FloatBarrier

\end{document}